\documentclass[nobalancelastpage,twocolumn,prl]{revtex4-1}
\usepackage{amsmath, mathtools}
\newcommand{\ket}[1]{\vert #1 \rangle}

\begin{document}

\title{A simple mechanism for unstable degeneracies in local Hamiltonians}

%

\author{Jos\'e Garre Rubio}
\affiliation{ {\small Instituto de F\'isica Te\'orica, UAM/CSIC, C. Nicol\'as Cabrera 13-15, Cantoblanco, 28049 Madrid, Spain}}

\begin{abstract}
If a local Hamiltonian eigenstate is mapped to another state by local operators commuting with the Hamiltonian terms, the latter is also an eigenstate. 
This basic observation implies a no-go result for both being a unique ground state and having a degeneracy protected against local perturbations.


\end{abstract}

\maketitle
The stabilizing mechanisms of ground spaces of local gapped Hamiltonians are key in quantum many-body physics, as global symmetries \cite{LandauVol5} and topological order \cite{Kitaev03}, and thereof understanding unstable degeneracies. Ref.\cite{gioia2024} showed that the W-state is not the unique ground state of any local Hamiltonian --even gapless and no frustration free (FF)-- and that such degeneracy is unstable. Here, we show a simple mechanism that enforces degenerate yet unstable energy levels covering the W-state result for FF Hamiltonians.

Let be $\ket{\psi}$ and $\ket{\phi}$ orthogonal states that are mapped by a set of finite-range operators $\{ O_\alpha\}_{\alpha \in \mathcal{S}}$ as
\begin{equation}\label{loconverti}
O_\alpha \ket{\psi} =\ket{\phi}, \  {\rm for \ all} \ \alpha \in \mathcal{S}\ .
\end{equation}
Suppose that $\ket{\psi}$ is an eigenvector of each term in a finite-range Hamiltonian $H=\sum h_j$, i.e., $h_j\ket{\psi} = e_j \ket{\psi}$ for all $j$, and that for each $h_j$ there exists an operator $O_\alpha$ such that $[h_j, O_\alpha] = 0$. Then, $\ket{\phi}$ is also an eigenvector of $h_j$ with the same eigenvalue $e_j$, since
$h_j \ket{\phi} = h_j O_\alpha \ket{\psi} = O_\alpha h_j  \ket{\psi} = e_j O_\alpha \ket{\psi} = e_j \ket{\phi}$.

Some comments are in order:
(1) The states $\ket{\psi}, \ket{\phi}$ are neither symmetry broken nor topological ordered states since the operators $O_\alpha$ are finite-range, excluding global symmetry operators and logical ones. (2) The permutation of $\ket{\psi} \leftrightarrow \ket{\phi}$ is only guaranteed to be local for $\ket{\psi} \rightarrow \ket{\phi}$: $\{O_\alpha\}$ may be non-invertible so $\ket{\phi} \rightarrow \ket{\psi}$ would  require non-LOCC operations. (3) For FF Hamiltonians satisfying $h_j\ket{\psi} = 0$, this implies that $\ket{\psi}$ cannot be the unique ground state. In particular, this is the case of the W-state $\ket{W} = \frac{1}{\sqrt{N}} \sum_i \sigma^+_i \ket{0}^{\otimes N}$, and any of its Dicke-state generalizations, since $$(N^{1/2} \sigma^-_i) \ket{W} = \ket{0}^{\otimes N} \ {\rm for \ any \ site} \ i,$$
so that either $\sigma^-_i$ or $\sigma^-_{i+\ell}$ commutes with all $\{h_j\}$ if $\ell\geq {\rm range}\{h_j\}$. So that $\ket{W}$ always appears together with $\ket{0}^{\otimes N}$ as ground states of FF Hamiltonians. (4) The W-example also shows that $[h_j, O_\alpha] = 0$ is satisfied easily when the supports of $h_j$ and $O_\alpha$ do not overlap.

Hastings showed \cite{Hastings06} that a finite-range gapped Hamiltonian $H=\sum_i \tilde{h}_i$ with ground state $\ket{\psi}$ is equivalent to another one $H=\sum_i {h}_i$ which ($i$) is almost FF: $|{h}_j \ket{\psi}|\leq |h_j| e^{-\ell}$, where $\ell$ can be picked as big as desired and ($ii$) is exponentially decaying: $|[h_j, O]| \leq |h_j| {\cdot} |O| e^{-d(h_j,O)/c}$, where $c$ depends on the microscopic details of $h_j$. Then, $\ket{\phi}$ in Eq.~\eqref{loconverti} is also close to a ground state:
\begin{equation}
| h_j \ket{\phi} | = |h_j O_\alpha\ket{\psi}| \leq |O_\alpha|{\cdot}|h_j|( e^{-\ell} + e^{-d(j,O_\alpha)/c}) \ ,
\end{equation}
provided that $d(j,O_\alpha)$ can be chosen arbitrary big. This extent our result to ground state degeneracies of non-necessarily FF Hamiltonians.


The degeneracies created in ground states of Hamiltonian given by Eq.~\eqref{loconverti} are unstable. For a local perturbation $H_\lambda = H_0 - \lambda \sum_j V_j$ that does not close the gap when $\lambda$ is small, a quasiadiabatic continuation follows \cite{Hastings05}: there is a unitary $U_\lambda$ that evolves the ground states $\ket{\psi}_\lambda=U_\lambda \ket{\psi}$ and preserves locality. To show the instability we compute:
\begin{align}
\langle H_\lambda \rangle_{\psi_\lambda} - \langle H_\lambda \rangle_{\phi_\lambda} & =
\langle H_0 \rangle_{\psi_\lambda} - \langle H_0 \rangle_{\phi_\lambda} \label{term1}
\\
&  -\lambda {\sum \mathop{}_{\mkern-5mu j}} \langle \tilde{V}_j-O^\dagger_{\alpha_j} O_{\alpha_j} \otimes \tilde{V}_j  \rangle_{\psi}\ \label{term2} 
\\ & = O(\lambda) \ ,
\end{align}
where $\tilde{V}_j = U^\dagger_\lambda V_j U^\dagger_\lambda$ is local and we have assumed that for every $j$ there is $\alpha_j \in \mathcal{S}$ such that $[\tilde{V}_j, O_{\alpha_j}] = 0$. Besides the first term --\eqref{term1}-- is expected to contribute exponentially small in $\lambda$, the second term --\eqref{term2}-- is linear in $\lambda$ times a contribution $ \sum_j\langle \tilde{V}_j  \rangle  (1-| O^\dagger_{\alpha} O_{\alpha} | f(d))$ which is $O(1)$ in $\lambda$, where the function $f$ encodes the correlation decay of $\psi$ and $d= max_j\{d(j,O_{\alpha_j})\}$. This linear growth in $\lambda$  discard a stable degeneracy, i.e. exponentially small separation in $\lambda$ between the ground state energies.


We emphasize that the results shown in this work do not require translation invariance and are valid for any physical dimension. Similar results can also be derived for the steady (mixed) states of frustration-free local Lindbladians mapped by local quantum channels.

{\it Acknowledgments.--} I thank A. Molnar and A. Turzillo, A. Franco-Rubio and A.M. Kubicki for enlightening discussions and the Erwin Schrödinger Institute (ESI), where this paper took shape. The author is funded by the FWF Erwin Schrödinger Grant DOI 10.55776/J4796.

\bibliography{bibliography}
\end{document}